\documentclass[useAMS, usenatbib]{mn2e}
\usepackage{xspace}
\usepackage{amsmath}
\usepackage{epsfig}
\usepackage{array}

\usepackage{defs}

\pdfoutput=1

\def\LW{Lyman-Werner\xspace}
\def\popIII{Pop~III\xspace}
\def\popII{Pop~II\xspace}

\def\ie{{\it i.e.}}
\def\eg{{\it e.g.}}
\def\ltsima{$\; \buildrel < \over \sim \;$}
\def\simlt{\lower.5ex\hbox{\ltsima}}
\def\gtsima{$\; \buildrel > \over \sim \;$}
\def\simgt{\lower.5ex\hbox{\gtsima}}

\def\ion#1#2{\text{#1\,\sc #2}}
\def\HI{{\ion{H}{i} }}
\def\HII{{\ion{H}{ii} }}

\def\H2{{H$_2$}}
\def\hide#1{}

\title[How I Wonder What You Are]{X-ray Twinkles and Pop~III Stars}

\author[M. Ricotti]{Massimo Ricotti$^{1,4}$\thanks{E-mail: ricotti@astro.umd.edu}\\
$^1$Department of Astronomy, University of Maryland, College Park, MD 20742, USA\\
}

\begin{document}

\maketitle

\begin{abstract}
Pop~III stars are typically massive stars of primordial composition
forming at the centers of the first collapsed dark matter structures.
Here we estimate the optimal X-ray emission in the early universe for
promoting the formation of Pop~III stars. This is important in
determining the number of dwarf galaxies formed before reionization
and their fossils in the local universe, as well as the number of
intermediate-mass seed black holes.

\noindent
A mean X-ray emission per source above the optimal level reduces the
number of Pop~III stars because of the increased Jeans mass of the
intergalactic medium (IGM), while a lower emission suppresses the
formation rate of H$_2$ preventing or delaying star formation in dark
matter minihalos above the Jeans mass.  The build up of the H$_2$
dissociating background is slower than the X-ray background due to the
shielding effect of resonant hydrogen Lyman lines. Hence, the nearly
unavoidable X-ray emission from supernova remnants of Pop~III stars is
sufficient to boost their number to few tens per comoving Mpc$^3$ by
redshift $z \sim 15$.

\noindent
We find that there is a critical X-ray to UV energy ratio emitted per
source that produces a universe where the number of \popIII stars is
largest: $400$ per comoving-Mpc$^3$. This critical ratio is very close
to the one provided by $20 - 40$~M$_\odot$ Pop~III stars exploding as
hypernovae.  High mass X-ray binaries in dwarf galaxies are far less
effective at increasing the number of \popIII stars than normal
supernova remnants, we thus conclude that supernovae drove the
formation of \popIII stars.
\end{abstract}

 
\section{Introduction}

The number of first stars (\popIII) per comoving volume that forms in
the early universe determines the level and homogeneity of metal
pre-enrichment of the intergalactic medium (IGM). Metal pre-enrichment
is important for modeling the formation of the first dwarf galaxies and
predict the number of pre-reionization fossils in the Local Group
\citep{RicottiG:05, BovillR:09}.  There are two main approaches widely
used for modeling the formation of the first dwarf galaxies in
cosmological simulations: (a) metal enrichment is calculated
self-consistently resolving the formation of Pop~III stars at $z>10$
in relatively small (1-4 cMpc$^3$) cosmological volumes
\citep{RicottiGSa:02, RicottiGS:08,Wise:08, Wise:14, Muratov:13a}; (b)
a metallicity floor (typically $Z \sim 10^{-3}$~Z$_\odot$) is
introduced everywhere in the IGM in order to run zoom simulations of
dwarf galaxies from high-redshift to the present \citep{Gnedin:10,
  Tassis:12, Christensen:12, Kuhlen:13, Hopkins:14, Thompson:14,
  Wheeler:15}.

The second method is not suited for capturing global feedback loops
that might affect the local metallicity floor and the intensity of the
radiation backgrounds, which are both important in determining the
fraction of dark matter halos that remain dark. However, the
self-consistent method in (a) also have limitations:
\begin{enumerate}
\item The gravitational potential of dark matter halos drives the
  collapse of proto-\popIII stars until the gas becomes
  self-gravitating at scales of a few AU \citep{Bromm1999,
    Abel2002}. Hence, in order to capture the formation of Pop~III
  stars is necessary to resolve the gravitational potential at the
  center of the minihalos of mass $10^5$~M$_\odot$ with at least
  several tens of particles. A dark matter resolution of about
  $100$~M$_\odot$ is required, setting a limit on the cosmological
  volume that can be simulated (e.g., $512^3$ simulation with
  $m_p=100$ M$_\odot$ has volume of 3~Mpc$^3$).
\item On the other hand, the small cosmological volume required to
  achieve the resolution necessary to resolve \popIII star formation
  in the smallest dark matter halos, prevents a self-consistent
  calculation of the radiation backgrounds that are important in
  determining the number of \popIII stars, especially in underdense
  regions where local feedback effects are sub dominant.
\end{enumerate}
One can choose to calculate the self-consistent backgrounds even
though the simulated volume is too small for numerical convergence, or
include a tabulated external background from analytical models. In the
first case, as soon as the very first Pop~III star is created, the
dissociating radiation background jumps from zero to a sufficiently
large value to destroy very rapidly all relic \H2 in the IGM
\citep[\eg,][]{RicottiGSb:02}. For the second choice, often a
tabulated background \citep[\eg,][]{HaardtMadau:12} is adopted. But in
the case of the formation of the Pop~III stars at $z=40-10$, the use of
backgrounds derived from observations at $z<10$ is not
justified. Hence, both choices are not satisfactory.

In this paper we use simple analytical calculations to estimate
self-consistently the number of \popIII stars in the early universe
and the radiation background they produce during their short life on
the main sequence and by their SN remnants. We also consider other
sources of X-rays: accreting intermediated mass black holes (IMBHs),
high mass X-ray binaries (HMXRBs) \citep{Xu:14, Jeon:14, Jeon:15}, and
miniquasars. We derive what is the level of X-ray emissivity that
maximizes the number of \popIII stars forming at $z \simgt 10-15$. The
basic idea is simple as noted by several authors before
\citep{Oh2001,Venkatesan2001, Machacek2003,RicottiO:04}.  An X-ray
background can both suppress \popIII star formation in the smallest
minihalos due to IGM heating (increasing the Jeans mass in the IGM)
and promote \popIII star formation by increasing the gas electron
fraction in gas collapsed into minihalos and thus promoting H$_2$
formation via the catalyst $H^-$.

The number of \popIII stars depends on the minimum dark matter halo
mass, $M_{\rm cr}$, in which a \popIII star can form as a function of
redshift; the smaller the critical mass the more numerous the \popIII
stars. However, $M_{\rm cr}$ depends on the X-ray background, as explained
above, and the H$_2$ dissociating background (UV in the Lyman-Werner
bands); since \popIII stars are responsible for producing the
dissociating and X-ray backgrounds, a feedback loop is in play.

The model is presented in \S~\ref{sec:model} and the results in
\S~\ref{sec:res}. The discussion is in \S~\ref{sec:disc}, and
summary and conclusions are in \S~\ref{sec:sum}.  We use Planck
cosmology $(\Omega_m, \Omega_\lambda, \Omega_b, h, n_s,
\sigma_8)=$(0.308, 0.692, 0.0482, 0.678, 0.968, 0.829)
\citep{Planck2015}.

\section{Model}\label{sec:model}

\begin{figure}
\includegraphics[width=9cm]{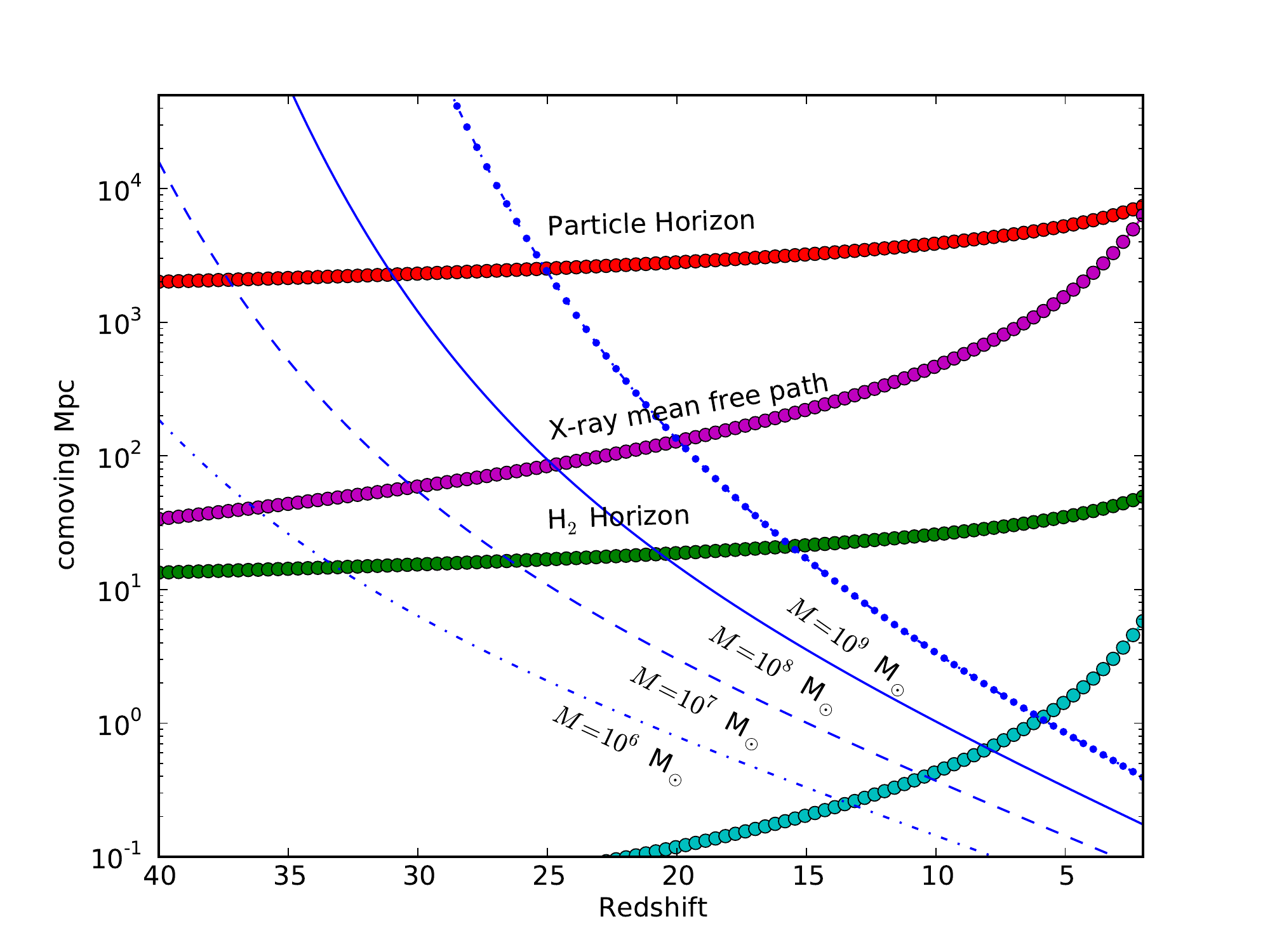}
\caption{\label{fig:mfp} Mean free path of H$_2$ dissociating
  radiation (green), hard UV at 54.4 eV (cyan), and soft X-rays at 0.5
  keV (magenta) in comparison to the particle horizon (red) and the
  mean distance between halos of mass $10^6, 10^7, 10^8$ and
  $10^9$~M$_\odot$ as indicated by the labels.}
\end{figure}

The first minihalos that form stars are rare objects in which massive
stars and their supernova remnants emit \H2 dissociating radiation in
the Lyman-Werner band and hard-UV/soft X-ray radiation (0.2-2 keV),
which have long mean free paths.  The mean free path of UV hydrogen
ionizing radiation is shorter than the distance between \popIII stars
until redshift $z \sim 10-15$ (neglecting clustering effects) when
cosmic \HII regions overlap and the UV ionizing background starts
dominating the reionization process. In our model we consider only
global backgrounds with mean free path longer than the distance
between \popIII stars: the \H2 dissociating radiation and a soft X-ray
background.  In particular, the inspiration for this paper came from
the realization that the mean free path of X-rays is longer than that
of the \H2 dissociation radiation that is shielded by resonant Lyman
lines in the mostly neutral IGM at $z>10$ \citep{Haiman2000,
  RicottiGS:01}. The mean free path of dissociating radiation, hard UV
and X-rays are shown in Figure~\ref{fig:mfp} as a function of
redshift, in comparison to the particle horizon and the mean distance
between halos of mass $M_{dm}$.

The formation of the first stars in a minihalo is only possible if two
conditions are met: the minihalo mass is larger than the Jeans mass
(which is $\sim 2\times 10^8~{\rm M}_\odot
(T_{\rm igm}/10^4~K)^{3/2}[(1+z)/10]^{-3/2}$) in the IGM, and that the gas
can cool sufficiently quickly.  It is well known that feedback from
X-rays can have both a positive or a negative effect on the formation
of \popIII stars: intense X-ray emissivity has negative feedback
because it heats the IGM, thereby increasing the critical dark matter
halo mass that is able to gravitationally attract and condense the IGM
gas, thus suppressing \popIII star formation in minihalos. A
negligible X-ray emissivity also has a negative impact on \popIII star
formation because, although gas can collapse into minihalos, it cannot
cool to the molecular form necessary to form stars. \H2 formation is
promoted by the reaction $H+H^- \rightarrow H_2+e^-$, where the
formation of the catalyst $H^-$ is maximized for electron fraction
$x_e \sim 0.1-0.5$ \citep{RicottiGS:01}. A weak level of partial
ionization by soft X-rays is accordingly ideal to promote H$_2$
formation and cooling because for $x_e<0.1$ the photoelectrons deposit
most of their energy into secondary ionizations rather than
heat. Therefore in this regime the effect of X-rays is to enhance the
ionization fraction of the IGM promoting H$^-$ and H$_2$ formation and
cooling, rather than increasing its temperature and Jeans mass that
instead would suppress \popIII star formation in small mass minihalos.

In summary, we expect that a weak X-ray emissivity, for instance
provided by the first SN remnants, may have the maximum effect at
boosting the number of \popIII stars and thus provide metal
pre-enrichment and beneficial ionizing radiation to continue star
formation in small mass halos formed before reionization.

In order to solve the problem analytically, we make the following assumptions:
\begin{enumerate}
\item[a.)] We only consider the formation of a single star per
  minihalo. Simulations have shown that \popIII stars may form in
  binaries or small multiple systems with a probability of roughly
  20\% \citep{Turk2009, StacyGB:10}. This scenario can also be
  considered in our model simply modeling the X-ray emission from
  possible HMXRB in addition to the total emission from the stars. We
  will touch on this in \S~\ref{sec:disc}.
\item[b.)] We assume that all stars are formed from gas of fully
  primordial composition. Thus, we neglect the contribution to the
  radiation backgrounds from \popII stars forming in minihalos
  externally enriched by powerful supernovae to a critical metallicity
  prior to collapse \citep{BrommF:01}. However, we include the
  contribution to the backgrounds from dark matter halos with masses
  $>10^8$~M$_\odot$, which are unaffected by reionization feedback and
  can cool emitting Lyman-$\alpha$ radiation.
\end{enumerate}
We model the fraction of bolometric energy per source emitted in the
X-ray band with the parameter $K_X$. Therefore the limit $K_X=0$
represent no X-ray emission. The remaining of section describes in detail the analytical
model with included physical processes and free parameters.

\subsection{Emissivity and backgrounds}\label{sec:emiss}
 
We make the simplifying assumption 
\begin{equation}
n_{pop3}(z) = n_{halo}(M>M_{\rm cr},z),
\end{equation}
where $n_{pop3}(z)$ and $n_{halo}(M,z)$ are the number density of \popIII
stars and dark matter minihalos of mass $M>M_{\rm cr}$, respectively. Basically we assume
that halos with mass $M<M_{\rm cr}$ remain dark while all halos with mass
$M>M_{\rm cr}$ host a \popIII star\footnote{ When a large fraction of dark
  matter halos fail to host galaxies, we expect a large scatter in the
  mass-to-light ratio due to local environmental effects. Here we
  neglect these local variations and we instead focus on how the
  global radiation backgrounds affect star formation in isolated
  minihalos.}. We derive $M_{\rm cr}(z)$ iteratively at each redshift
as explained in \S~\ref{ssec:loop}.

Given the number density of dark matter halos $n_{halo}$ (see Appendix~A for a
fitting formula based on Press-Schechter formalism), the bolometric
emissivity is
\begin{equation}
\epsilon(M_{\rm cr},z)=n_{halo}(M>M_{\rm cr},z)L_{*}^\prime f_{\rm duty} 
% + n_{halo}(10^8,z)L_*f_{duty},
\end{equation}
where $f_{\rm duty}=t_{\rm on}^\prime/t_H(z)$ is the fraction of Hubble
time the sources are emitting radiation at luminosity $L_{*}^\prime$
and $t_{on}^\prime$ is the typical lifetime during which a star emits
radiation before leaving the main sequence and exploding as a SN or
Hypernova. Here we adopt parameters $L_*^\prime=10^{40}$~ergs/s and
$t_{\rm on}^\prime=2$~Myr typical of a 100~M$_\odot$ \popIII star, or
a total emitted energy per source $E_{pop3}=L*^\prime t_{on}^\prime=6
\times 10^{53}$~ergs. In general, the values of $t_{\rm on}$ and $L_{*}$
depend on the initial mass function (IMF) of \popIII stars. Very massive stars
($>100$~M$_\odot$) emit near the Eddington limit, thus $L_*\approx
10^{40}~{\rm ergs/s}(M_*/100~M_\odot)$. Since we have assumed a fixed
$L_{*}^\prime$ and $t_{on}^\prime$, typical of 100~M$_\odot$ \popIII
stars, we can model an arbitrary IMF by adjusting two of the three
free parameters in our model: $K_{LW} \equiv L_{LW}t_{on}/E_{pop3}$,
the mean energy emitted per source (or, equivalently, per minihalo) in
the \LW bands, and $K_X \equiv L_X t_{on}/E_{pop3}$, the mean energy
per source (minihalo) in the soft X-ray band. Thus, the emissivity in
a given frequency band is $\epsilon_{band}(M,z) = K_{band}
\epsilon(M,z)$. We refer to \S~\ref{sec:disc} for details on the
typical values of $K_{LW}$ and $K_X$ for different IMFs of \popIII
stars and X-ray sources (\eg, SNe, Hypernovae, HMXRBs).  For
reference, the free parameters in our model are listed in
Table~\ref{tab:free}.

\begin{table}
\centering
\caption{List of free parameters in the model.}
\label{tab:free}
\begin{tabular}{l m{5cm}} % four columns, alignment for each
\hline
Parameter & Explanation \\
\hline
$K_{LW}\equiv \frac{E_{LW}}{6\times 10^{53}~{\rm ergs}}$ & $E_{LW}$ is the mean energy emitted from \popIII stars per minihalo in the Lyman-Werner bands.\\
$K_X \equiv \frac{E_X}{6\times 10^{53}~{\rm ergs}}$   & $E_X$ is the mean energy emitted per minihalo in the X-ray bands ($0.2-2$~keV). Note that $\beta_X \equiv K_X/K_{LW}$ is a derived parameter.\\
$T_0\equiv \frac{\Gamma_\HI}{k_B\zeta_\HI}$   & Characteristic spectral temperature of the sources, proportional to the ratio of the photo-heating to photo-ionization rate. It depends on the source spectrum modified for ISM absorption. \\
\hline
\end{tabular}
\end{table}
In addition, we include the rise of galaxies forming in halos with
masses $>10^8$~M$_\odot$, that are not subject to either thermal or
cooling feedback (since their virial temperature is $>10^4$~K, the gas
is collisionally ionized and can cool by Lyman-$\alpha$
emission). Note that $M_{\rm cr} < 10^8$~M$_\odot$ in all cases. We
thus add to the emissivity in the X-ray and \LW bands the term
\begin{equation}
\epsilon_{band}(>10^8~M_\odot,z)=K_{UV}n_{halo}(10^8~M_\odot,z)L_*^\prime f_{duty},
\end{equation}
where $K_{UV}=1$ is chosen to reheat and reionize the IGM at $z \sim
5-6$.  These ``Lyman-$\alpha$ cooling'' galaxies are thought to be the
primary agents of hydrogen reionization \citep{RicottiGSb:02,
  Robertson:15} but have negligible effect on the IGM, the number of
\popIII stars and the radiation backgrounds at $z>15$. The decline of
the number of \popIII stars and the radiation backgrounds they produce observed
in our models at $z \simlt 10$ is due to the rise of this more massive
population of galaxies.

Given the specific emissivity, the mean specific intensity of the
background $J_\nu \equiv J_0 g_\nu$ is
\begin{equation}
4\pi J_\nu=\min{(\lambda_\nu, ct_H)} \epsilon_\nu(M_{\rm cr},z),\label{eq:jnu}
\end{equation}
where $\lambda_\nu=(\sigma_\nu n_{HI})^{-1}$ is the photon mean free
path. When the mean free path is smaller than the particle horizon
(for instance for photon energies $< 2$~keV in the mostly neutral IGM
at $z>10$), we have $J_\nu \propto \epsilon_\nu/n_{H}
\sigma_\nu$. Thus, in this case the ionization and heating rates [see
  Eqs.~(\ref{eq:ion})-(\ref{eq:heat})] become independent of the
ionization cross section because all photons are eventually absorbed
before being significantly redshifted.

\subsection{Electron fraction and temperature of the IGM}

In order to know whether a minihalo is able to form a \popIII star we
need to determine whether the gas in the IGM can condense into the
gravitational potential of minihalos of mass $M$. We thus need to
calculate the Jeans mass of the IGM and therefore its temperature
evolution. A gas of primordial composition in minihalos of mass
$<10^8$~M$_\odot$ can cool and form stars only if a sufficient amount
of H$_2$ is formed in the minihalo. As explained above, the formation
of molecular hydrogen is catalyzed by the ion $H^-$ formed by the
reaction $H+e^- \rightarrow H^- + \gamma$, whose rate depends on the
ionization fraction of the gas. Both temperature and ionization
fraction are determined by the radiation backgrounds, that in turn are
proportional to the number of minihalos hosting a \popIII star,
establishing a feedback loop.

The evolution of the electron fraction and temperature of the IGM
depend on the ionizing radiation background by the following
equations:
\begin{align}
\frac{d x_e^{\rm igm}}{dt} &= f_i \zeta_{HI},\\
k_B\frac{d T^{\rm igm}}{dt} &= f_h \Gamma_{HI}. 
\end{align}
where,
\begin{align}
\zeta_{HI} &=4\pi J_0\int \frac{d\nu}{h\nu}g_\nu \sigma_\nu,\label{eq:ion}\\
\Gamma_{HI} &=4\pi J_0\int d\nu g_\nu \sigma_\nu \equiv k_B T_0 \zeta_{HI},\label{eq:heat} 
\end{align}
are the ionization and heating rates, respectively. The characteristic
temperature $T_0$ depends only on the spectrum $g_\nu$ of the source
modified for ISM absorption, and is the third free parameter in the
model.  Integrating the equations above we obtain:
\begin{align}
x_e^{\rm igm} &=f_i \frac{\Gamma_{HI}t_H}{k_B T_0},\\
k_B T^{\rm igm} &=f_h \Gamma_{HI} t_H,
\end{align}
where $\Gamma_{HI}=K_X \epsilon/n_{HI}$. Thus, the temperature of the IGM
can be easily determined from the electron fraction. In summary:
\begin{align}
x_e^{\rm igm} (M_{\rm cr},z) &=f_i \frac{K_X \epsilon(M_{\rm cr},z)}{k_B T_0}\frac{t_H}{n_H},\\
T^{\rm igm} (M_{\rm cr},z) &= T_0 x_e^{\rm igm} \frac{f_h}{f_i},
\end{align}
where the functions $f_h(x_e), f_i(x_e)$, given in appendix~B, account
for the effect of secondary ionization from fast photoelectrons
\citep{ShullVanS:85, ValdesFerrara:08}. A minimum $x_e^{\rm igm}=10^{-4}$,
due to the residual ionization after recombination
\citep[\eg,][]{Tegmark:97}, is set as a ionization
floor\footnote{There are no other processes setting a floor value for
  the electron fraction at high-z other than exotic sources like
  primordial BHs \citep{RicottiOM:08} or decaying dark matter
  \citep{MapelliF:2006}. At lower redshifts escaping UV radiation may
  produce short lived bubbles of ionized gas that after recombining
  leave behind long-lived relic \HII regions of partially ionized
  gas \citep{HartleyRicotti2016}. Inside those relic \HII regions
  \popIII star formation might be enhanced, but we have neglected this
  effect.}. Similarly, the temperature floor of the IGM in absence of
heating sources is given by the cosmic adiabatic expansion after
radiation-gas thermal decoupling at $z \sim 100$: $T^{\rm igm}=300~K
[(1+z)/100]^{2}$.

The X-ray and \LW backgrounds can also be derived as a function of
$T^{\rm igm}$. For the \LW band the mean free path in the mostly neutral IGM is about
150 times smaller than the particle horizon due to the shielding
effect of resonant Lyman-$\alpha$ lines \citep[see Fig.~12
  in,][]{RicottiGS:01}. Thus, from Equation~(\ref{eq:jnu}) and $K_X
\epsilon=\Gamma_{HI}n_H = k_B T^{\rm igm}/(f_h t_H)$, we get:
\begin{equation}
4\pi J_{LW}=K_{LW} \epsilon(M_{\rm cr},z) \frac{ct_H}{150}=\frac{K_{LW}}{K_X} \frac{c n_H(z)}{150 f_h} k_B T^{\rm igm}.
\end{equation}

\subsection{Formation of \popIII stars and feedback loop}\label{ssec:loop}

The formation of \popIII stars in a minihalo of mass $M$ is only possible if
these two conditions are met:
\begin{enumerate}
\item {\bf $T_{\rm vir}(M,z) > T^{\rm igm}(M_{\rm cr},z)$:} the dark matter halo mass is larger than Jeans mass of the IGM. This condition is satisfied for dark matter halo masses $M>M_{\rm cr}^{i}$ . 
\item {\bf $t_{\rm cool}(M,z) < t_H(z)$:} the virialized gas inside dark matter halos of mass $M$ cools in less than the Hubble time. This condition is satisfied for dark matter halo masses $M>M_{\rm cr}^{ii}$ . 
\end{enumerate}
For a given temperature of the IGM, $T^{\rm igm}$, only halos with virial
temperature $T_{\rm vir}>T^{\rm igm}$ can condense gas from the IGM. A gas at
the virial temperature in hydrostatic equilibrium in the potential of
a dark matter halo with virial radius $r_{\rm vir}$, has a gas core radius
$r_c \sim (0.22/c) r_{\rm vir}$ with gas overdensity $\delta_c \sim 2000$
\citep{Ricotti:09}. Here we have assumed halo concentration parameter
$c=5$.  A hot IGM increases the minimum dark matter halo mass that is
able to form stars, $M_{\rm cr}^{i}$, thus reduces the number of X-ray
sources and the X-ray background.  But a lower X-ray background
translates into lower IGM temperature. Clearly, this is the first
feedback loop in play.
  
The second condition is:
\begin{equation}
t_{\rm cool}=\frac{k_B T_{\rm vir}}{\delta_c n_H x_{H_2} \Lambda(T_{\rm vir})}<t_H(z),
\end{equation}
where $\Lambda(T)$ [erg cm$^3$ s$^{-1}$], given in Appendix~B,
is from \citep{Galli:98}.  Condition (ii) sets a minimum $x_{H_2}$ for
collapse. The \H2 abundance depends on the formation of $H^-$ and is
regulated by the gas temperature in virialized halos,
$T_{\rm vir}(M,z)$, as well as the dissociating radiation background and the
electron fraction in the halo according to the equation:
\begin{equation}
x_{H_2}(M,z) = n_e \frac{k_1(T_{\rm vir})}{k_2(G_{LW})}\left(1+\frac{N_{H_2}}{10^{14} cm^{-2}}\right)^{0.75},
\end{equation}
where $G_{LW} \equiv J_{LW}/(1.6 \times 10^{-3}~{\rm erg/s/cm}^2)$.
The H$^-$ formation rate, $k_1$, and the H$_2$ dissociation rate, $k_2$,
are given in Appendix~B. We have also included the effect of H$_2$
self-shielding, that is important for H$_2$ column densities $N_{H_2}
\equiv r_{c} n_{H_2}>10^{14}$~cm$^{-2}$, where $r_c$ is the core
radius of the gas in hydrostatic equilibrium in the dark matter halo.

Inside dark matter halos (with gas overdensity $\delta_c$) the electron
fraction differs from the value in the IGM because of the larger
recombination rate. Thus, the equilibrium electron fraction inside
dark matter halos is:
\begin{align}
n_e & =\delta_c n_H \sqrt{x_e^{\rm igm} R}\\
R &=\min(x_e, t_{\rm rec}/t_H).
\end{align}
Thus, condition (ii) sets a feedback loop that regulates the number of
\popIII stars as follows: lowering the minimum halo mass hosting
\popIII stars $M_{\rm cr}^{ii}$ with respect to the feedback regulated
value increases the number of sources producing a higher electron
fraction but also a higher dissociating background. The \H2 formation
rate and the gas temperature (equal to the halo virial temperature)
are both reduced resulting in a cooling time longer than the Hubble
time. Thus $M_{\rm cr}^{ii}$ needs to increase to satisfy condition (ii).

In order to satisfy both conditions (i) and (ii) the critical mass
must be
\begin{equation}
M_{\rm cr}(z) = \max{[M_{\rm cr}^{i}(z), M_{\rm cr}^{ii}(z)]}.
\end{equation}
When the critical mass is set by the cooling condition (\ie, when
$M_{\rm cr}=M_{\rm cr}^{ii}$ and $M_{\rm cr}^{i}<M_{\rm cr}^{ii}$), increasing the
X-ray emissivity $K_X$ produces higher electron fraction and \H2
fractions in dark matter halos, reducing $M_{\rm cr}^{ii}$, thus
increasing the number of \popIII stars. At the same time increasing
$K_X$ increases the IGM temperature and the Jeans mass in the
IGM. Thus $M_{\rm cr}^i$ increases approaching $M_{\rm cr}^{ii}$.  The maximum
number of \popIII stars at any given redshift (for a fixed $K_{LW}$)
is obtained for a value of $K_X=K_X^{\rm cr}$ that makes the critical
masses from conditions (i) and (ii) equal to each other (\ie,
$M_{\rm cr}^i = M_{\rm cr}^{ii}$). For $K_X>K_X^{\rm cr}$ the critical mass is set
by the Jeans condition ($M_{\rm cr}=M_{\rm cr}^i$ and
$M_{\rm cr}^{ii}<M_{\rm cr}^{i}$), thus increasing $K_X$ leads to a higher
Jeans mass in the IGM, a higher $M_{\rm cr}$ and reduces the number of
\popIII stars.

\section{Results}\label{sec:res}

\begin{figure}
\includegraphics[width=8.5cm]{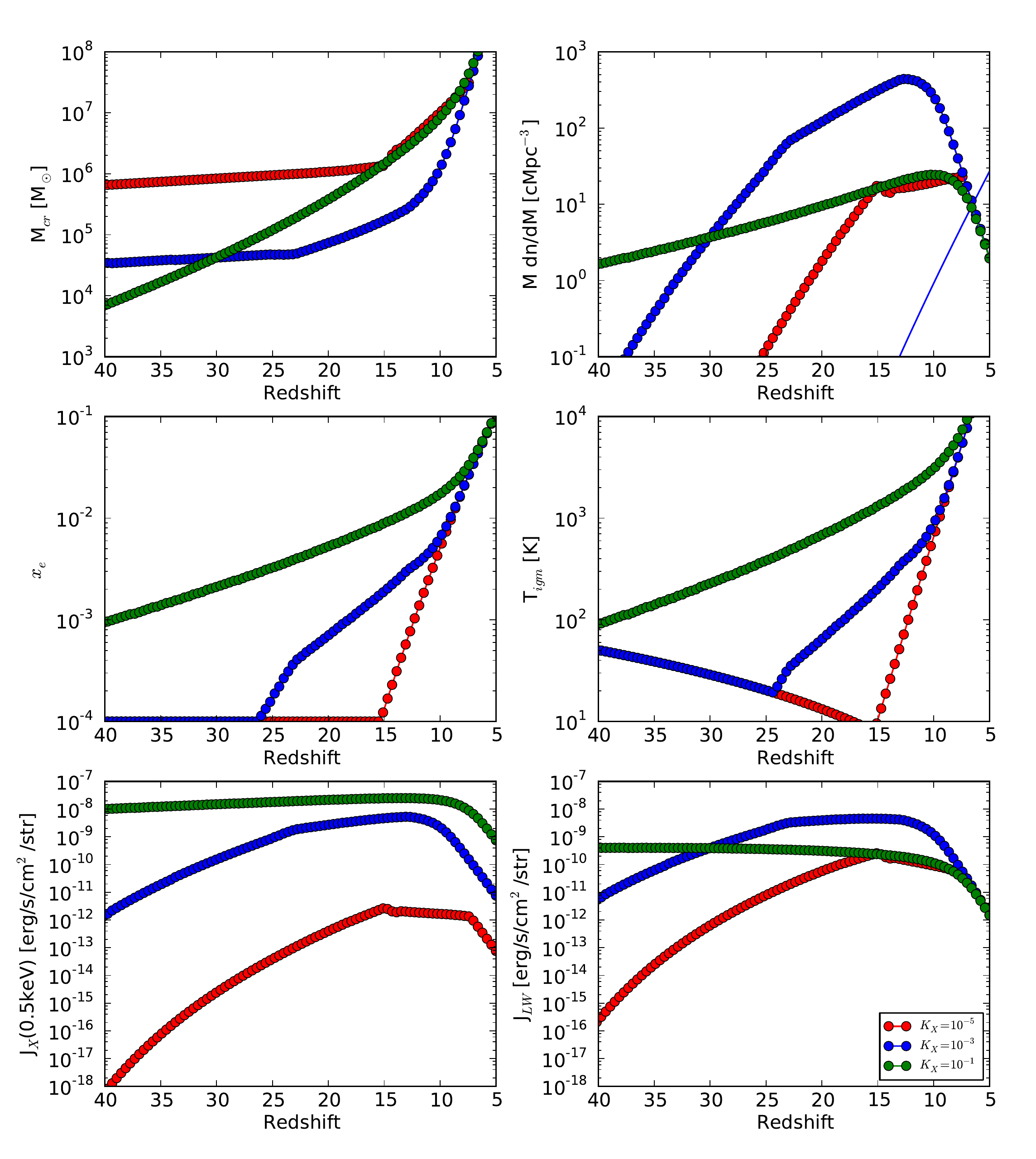}
\caption{\label{fig:runs} Simulations of \popIII star formation and
  its effect on the IGM for three different models differing for the
  X-ray emissivity of the sources: (Top left.) Evolution of the
  critical mass $M_{\rm cr}$ for \popIII star formation as a function of
  redshift, $z$. (Top right.) Number of \popIII stars per comoving Mpc
  as a function of $z$. The thin blue line in this panel shows the
  number of Lyman-$\alpha$ cooling halos ($M>10^8$~M$_\odot$). (Middle
  left.) Electron fraction $x_e^{igm}$, and (middle right) temperature
  of the IGM, $T^{igm}$, as a function of $z$. The bottom panels show
  the evolution of the X-ray background (bottom left) and \H2
  dissociating background (bottom right). The simulation parameters
  are $T_0=2 \times 10^5$~K, $K_{LW}=10^{-2}$ and $K_X=10^{-5}$ (red),
  $10^{-3}$ (blue), and $10^{-1}$ (green). See the text for the
  meaning of the free parameters $T_0$ and $K_{LW}$.}
\end{figure}
The initial mass function for the first stars and the nature of their
relics are very uncertain, so we start by examining three cases that
illuminate the range of available parameter space. We will keep the
ultraviolet luminosity and spectrum of the sources fixed adopting
$K_{LW}=10^{-2}$ and $T_0=2 \times 10^5$~K (typical of $\sim
40$~M$_\odot$ \popIII stars) and vary $K_X$ considering the cases: i)
low X-ray energy per source (\eg, the baseline X-ray emission from
normal SN explosions); ii) moderate X-ray energy (\eg, if \popIII
stars explode as hypernovae); iii) strong X-ray energy per source
(\eg, produced by an hypothetical population of miniquasars or
accreting intermediate mass black holes).

In Figure~\ref{fig:runs} we show the results for the three sets of
runs. The top two panels show the evolution as a function of redshift
of the critical mass $M_{\rm cr}$ for \popIII formation and the number
of \popIII stars per comoving Mpc$^3$. The two panels in the middle show
the evolution as a function of redshift of the electron fraction,
$x_e^{\rm igm}$, and temperature, $T^{\rm igm}$, of the IGM. The bottom two
panels show the evolution of the soft X-ray background at $0.5$~keV,
$J_X$, and the H$_2$ dissociating background in the \LW bands,
$J_{LW}$.

For the low X-ray energy case ($K_X=10^{-5}$, red lines in
Figure~\ref{fig:runs}), the X-ray background has a small effect on the
evolution of the ionization faction ($x_e^{\rm igm}$) and temperature of
the IGM ($T^{\rm igm}$) until redshift $z \simlt 15$, when halos with mass
$>10^8$~M$_\odot$ start dominating the UV and X-ray emission. 
The Jeans mass in the IGM (\ie, the minimum mass that can collapse
gravitationally) is small but the gas collapsed into minihalos does
not cool and therefore cannot form stars because the H$_2$ cooling
time is longer than the Hubble time, which at $z=15$ is $272$~Myr. In
this model the minimum halo mass in which \popIII stars can form
($M_{\rm cr}$) is the largest of the three models, and the total number of
\popIII stars at $z \sim 15$ (and the radiation background they
produce) is the lowest among the three models.  In all the runs at $z<15$
Lyman-$\alpha$ cooling halos with $M>10^8$~M$_\odot$, which are not
subject to the feedback loop, start to dominate the radiation
emissivity. The reheating of the IGM and \LW background produced by
these halos start suppressing the formation of \popIII stars. However,
several simplifying assumptions in our model become increasingly
unrealistic at $z<10-15$. For instance, metal enrichment from powerful
winds in these halos is expected to suppress \popIII star formation.

\noindent
In the model with strong X-ray energy per source ($K_X=10^{-1}$, green
lines), the ionization fraction and temperature of the IGM rise
rapidly. The minimum mass of dark matter halos in which \popIII star
can form ($M_{\rm cr}$) coincides with the IGM Jeans mass, which increases
rapidly with time. Thus, the number of \popIII stars is relatively
large at high-redshift but does not increase significantly with
time. The intensity of the X-ray background $J_X$ is the highest with
respect to the three models and rather constant with redshift, while the
dissociating background $J_{LW}$ is the highest at high-z but becomes
lower than in the intermediate model with decreasing redshift.

\noindent
The model with moderate X-ray energy ($K_X=10^{-3}$, blue lines)
maximizes the number of \popIII stars per unit volume at $z \sim 15$.
The maximum number of \popIII stars is obtained when the critical halo
mass in which gas can cool in a Hubble time equals the Jeans mass of
the IGM. The background in the \LW band is the highest and the X-ray
background is also very close to the highest among the three models.
The maximum number of \popIII stars is reached around redshift $z \sim
15$, before the radiation by normal stars dominates the background.
\begin{figure*}
\includegraphics[width=9cm]{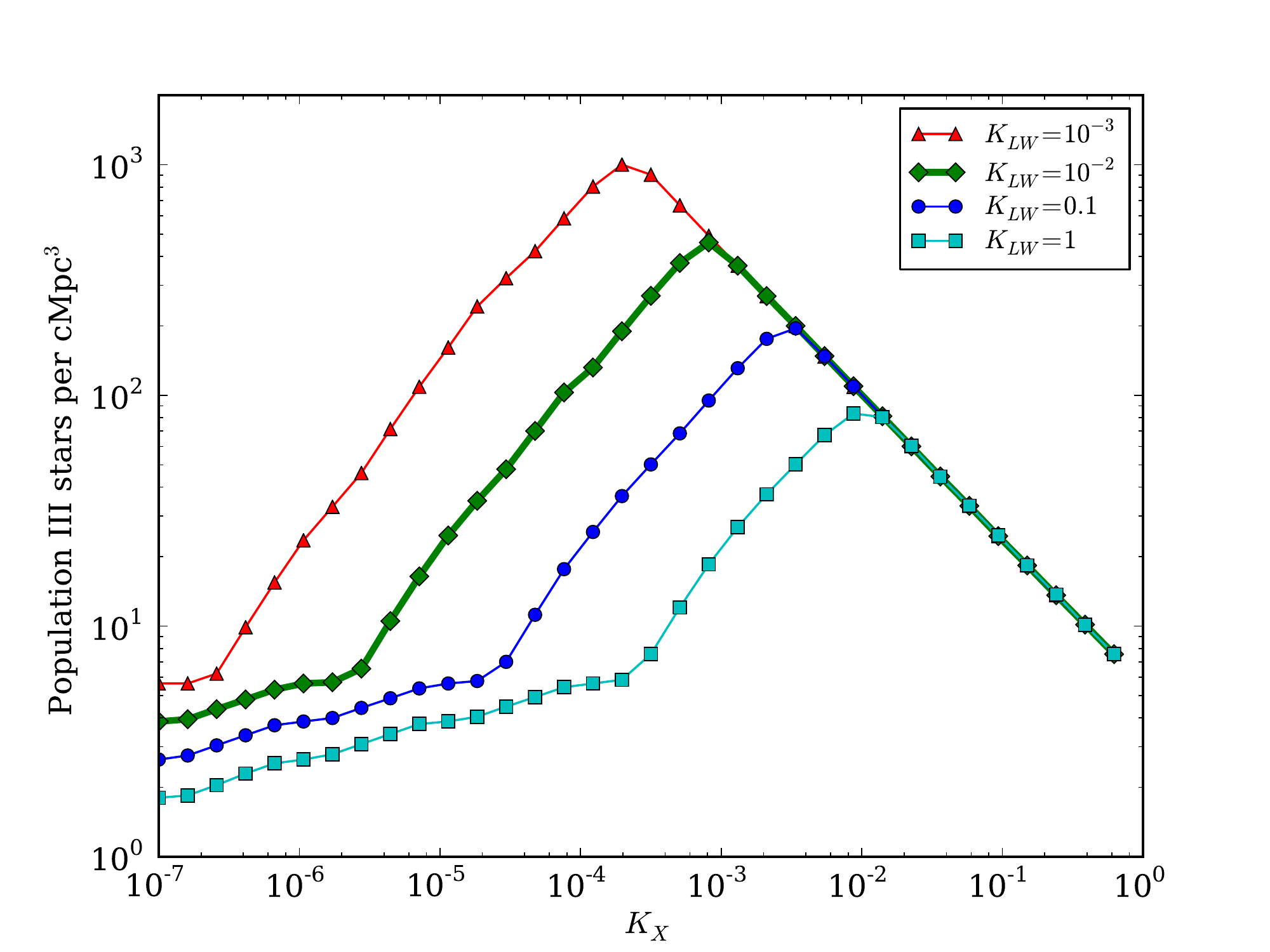}%
\includegraphics[width=9cm]{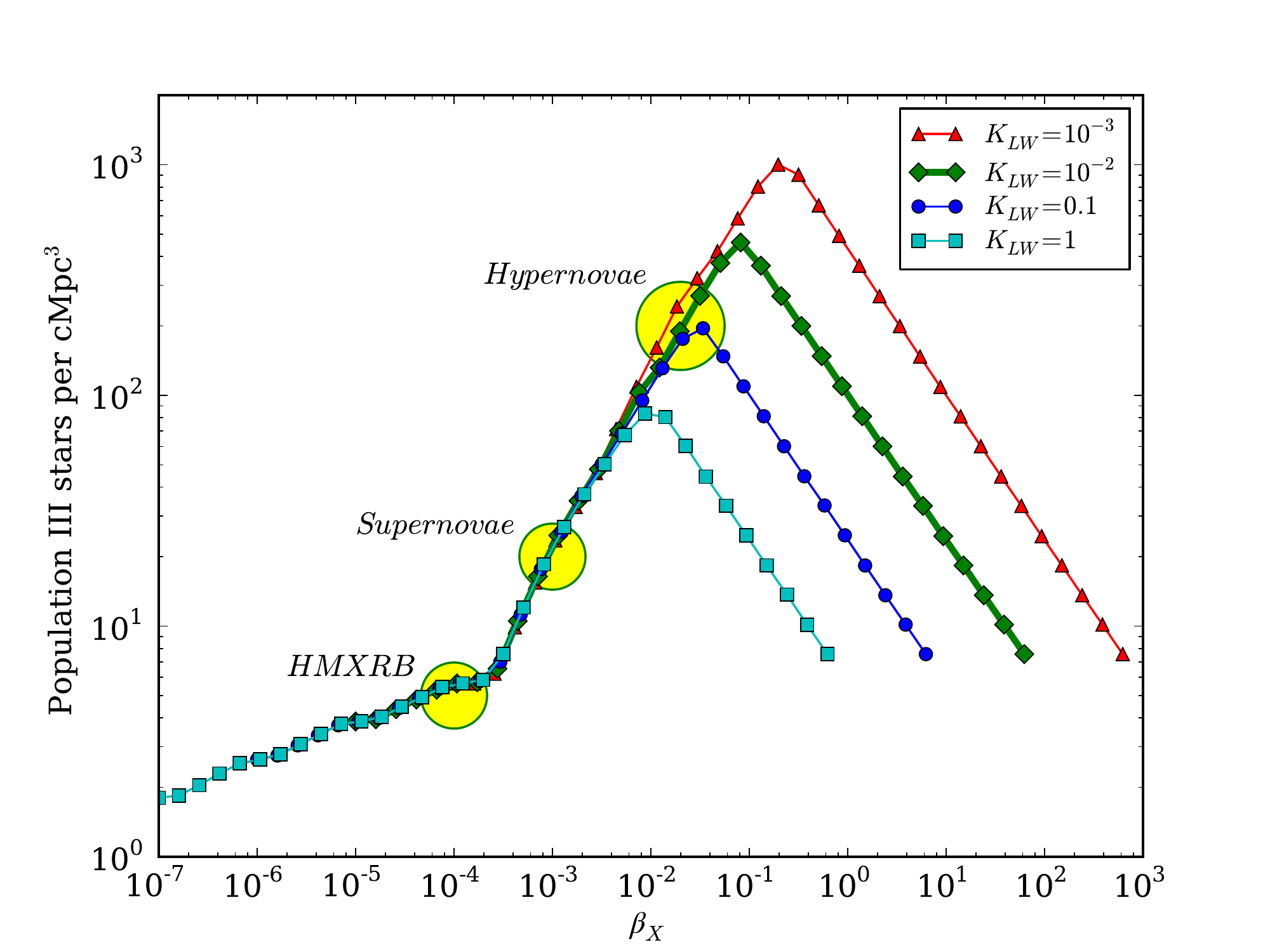}
\caption{\label{fig:params} (Left) Maximum number of \popIII stars per
  comoving Mpc$^3$ forming at a given redshift (around $z\sim 10-13$)
  as a function of $K_X$, for different values of $K_{LW}$, as shown
  in the legend.  The parameters $K_{LW}$ and $K_{X}$, are the mean UV
  (in the Lyman-Werner band) and soft X-ray energy emitted by each
  source in units of $6 \times 10^{53}$~ergs (roughly the bolometric
  energy emitted over its lifetime by a \popIII star of
  100~M$_\odot$). (Right) Maximum number of \popIII stars per comoving
  Mpc$^3$ (forming at a given redshift) as a function of $\beta_X
  \equiv K_X/K_{LW}$, for different values of $K_{LW}$. The yellow
  circles show approximate values of $\beta_X$ for different X-ray
  sources as indicated by the labels: high mass X-ray binaries
  (HMXRBs), supernovae and hypernovae.}
\end{figure*}

In Figure~\ref{fig:params} we show the maximum number of \popIII stars
per comoving Mpc$^3$ (reached at redshift $z^{max}$), as as a function
of $K_X$ (left panel), and as a function of $\beta_X \equiv
K_X/K_{LW}$ (right panel) for different values of $K_{LW}$ as shown in
the legend.  For low values of $K_X$, the maximum number of \popIII
stars, $n_{pop3}(z^{max})$, increases as $\beta_X$ increases (achieved
by either increasing $K_X$ or decreasing $K_{LW}$). This is the regime
where the number of \popIII stars is limited by the H$_2$ formation
rate and thus the ability of the gas collapsed inside minihalos to
cool in less than a Hubble time. For higher $K_X$ the number of
\popIII stars is independent of $K_{LW}$ and decreases with increasing
$K_X$. In this regime the number of \popIII stars is limited by the
heating of the IGM, which prevents gas from collapsing into minihalos.

The number of \popIII stars per cMpc$^3$ is approximated by a broken
power:
\begin{equation}
n_{pop3}=6~{\rm cMpc}^{-3}\times \\
\begin{cases}
\left(\frac{\beta_X}{2\times 10^{-4}}\right)^{0.15} ~\text{if $\beta_X < 2 \times 10^{-4}$}\\
\left(\frac{\beta_X}{2\times 10^{-4}}\right)^{0.73} ~\text{if $2\times 10^{-4}<\beta_X < \beta_X^{\rm cr}$}\\
K_X^{-0.73} ~\text{if $\beta_X> \beta_X^{\rm cr}$},
\end{cases}
\end{equation}
and reaches it maximum value
\begin{equation}
n_{pop3}^{max}=430~{\rm cMpc}^{-3} \left(\frac{K_{LW}}{10^{-2}}\right)^{-0.365},
\end{equation}
when the critical mass for cooling and the IGM Jeans mass coincide.
This happens for $\beta_X=\beta_X^{\rm cr}=0.01 K_{LW}^{-0.5}$ or
equivalently, $K_X=K_X^{\rm cr}=0.01 K_{LW}^{0.5}$.

For our fiducial choice $K_{LW}=10^{-2}$, typical of $40$~M$_\odot$
\popIII stars (see \S~\ref{sec:disc}), we find $\beta_X^{\rm cr}=0.1$
(or $K_X^{\rm cr}=10^{-3}$).  We will show in the next section that
this corresponds best to a population of $\sim 40$~M$_\odot$ stars
exploding as hypernovae (see yellow circles in the figure for the
parameters of different X-ray sources).  We also find that the
contribution of HMXRBs corresponds to a ratio $\beta_X=10^{-4}$ (see
\S~\ref{sec:disc}) and therefore they play no role in either
suppressing or enhancing the formation of \popIII star.

\section{Discussion}\label{sec:disc}

We have not discussed yet the type of sources that can provide the
level of X-ray and \LW emissivity able to boost or suppress the
formation of \popIII stars. Our results are given in terms of the
dimensionless energies emitted by each source in the X-ray and \LW
bands: $K_X$ and $K_{LW}$, respectively. The normalization we adopted
(see \S~\ref{sec:emiss}) is typical of $100$~M$_\odot$ \popIII stars:
$E_{pop3}=L_*^\prime t_{on}^\prime= 10^{40}~{\rm erg/s} \times 2~Myr = 6 \times
10^{53}$~erg.  Hence, as discussed before, the value of $K_{LW}\equiv
E_{LW}/(6 \times 10^{53}~{\rm ergs})$, depends on the IMF of \popIII
stars. We consider three representative cases \citep{Helgason:15}:
\begin{enumerate}
\item {\bf Very massive \popIII stars ($100 - 500$~M$_\odot$).} They
  emit close to the Eddington limit and a significant fraction of
  their bolometric luminosity is in the UV bands: the luminosity per
  solar mass in stars in the \LW bands is $2.7 \times 10^{37}$~erg/s and
  $t_{on}=2$~Myr, hence:
\[
K_{LW}=2.7\times 10^{-1} \left(\frac{M_*}{100~M_\odot}\right).
\]
\item {\bf Massive \popIII stars ($10-40$ M$_\odot$).} In this mass
  range the luminosity per solar mass is sub-Eddington: $L_{LW}=2.7 \times
  10^{36}$~erg/s and $t_{on}=2$~Myr, hence
\[
K_{LW}=10^{-2}\left(\frac{M_*}{40~M_\odot}\right).
\]
\item {\bf Metal poor stars with Kroupa IMF.} The luminosity per solar
  mass is sub-Eddington but the lifetime of the stars is longer:
  assuming masses between 0.1 and 100 M$_\odot$ with metallicity of
  Z=0.0004, the mean luminosity per star (which has mean mass of
  $0.64$~M$_\odot$) is $L_{LW}=0.64 \times 10^{36}$~erg/s, and
  $t_{on}=10$~Myr. Hence, still assuming one star per halo, we estimate:
%6.4e36/2e40
\[
K_{LW}=3.2 \times 10^{-4}.
\]
\end{enumerate}
In the following subsections we estimate the values of $K_X$ and
$\beta_X$ for different X-ray emitting sources, starting from SN
explosions and SN remnants of \popIII stars. To estimate $\beta_X$
we assume $K_{LW}$ in case (ii).

\subsection{Supernova remnants of \popIII stars}

The first SNe are a nearly unavoidable outcome of the formation of the
\popIII stars (unless all \popIII stars have masses $M_*>250$~M$_\odot$
and collapse into BHs without exploding \citep{HegerWoosley:02}). Thus,
SNe set a minimum floor for the X-ray emissivity. The assumption that
the X-ray and \LW emissivity are directly proportional to the number
of \popIII stars is clearly satisfied in this case.

Normal SNe with standard energy of the explosion,
$E_{SN}=10^{51}$~erg, emit mostly soft X-rays in band
$0.2-2$~keV. The X-ray luminosity from the hot gas in the SN cavity is
about $L_X=10^{37}$~erg/s, and it emits for about $t_{on}=10^4$~yr
before cooling. Thus, the total energy in the soft X-ray bands is
$E_{SNR}=3 \times 10^{48}$~erg \citep{Lopez:11}.  The supernova
explosion itself also emits in the X-ray bands with peak luminosity
$L_X=10^{42}$~erg/s lasting about $t_{on}=1$~month. Thus, the total
emitted energy in X-rays is comparable to the remnant's energy:
$E_{SN} \approx 2.6 \times 10^{48}$~erg.  The total energy emitted in
the soft X-ray band from each \popIII SN explosion (independently of
the \popIII mass as long as $M_*>8$~M$_\odot$) is $E_{X,
  tot}=E_{SN}+E_{SNR} \approx 6 \times 10^{48}$~erg, corresponding to
$K_X=10^{-5}$. Hence, for the fiducial $K_{LW}$ we find
\begin{equation}
\beta_X({\rm SNe})=10^{-3}\left(\frac{M_*}{40~M_\odot}\right)^{-1}.
\end{equation}

We speculate that if a fraction of \popIII stars explode as hypernovae
\citep{UmedaNomoto2003} or pair-instability SNe (PISNe), that are about
10 to 100 times more energetic than normal SNe, then $K_X \sim
10^{-3}-10^{-4}$ and
\begin{equation}
\beta_X({\rm Hypernovae}) \sim 10^{-1}-10^{-2},
\end{equation}
depending on the details of the emitted energy and the fraction of
\popIII stars that explode as hypernovae. Thus, an exciting insight
from this study is the following: if a non-negligible fraction of
\popIII stars end their lives as hypernovae, the soft X-rays from
their remnants and explosions is sufficient promote the formation of
\popIII stars to about 400 per cMpc$^3$, that is near the
maximum value found for any choice of $K_X$ and the assumed IMF of \popIII stars.

\subsection{High mass X-ray binaries (HMXRB)}

The X-ray emission does not come directly from \popIII stars in this
case but from normal dwarf galaxies with more than one star. Thus, we
have to make the additional assumption that the emission from normal
galaxies is proportional to the number of minihalos hosting a \popIII
star.

To estimate the X-ray emission from low metallicity HMXRB we use the
empirical relationship \citep{Fragos:13}:
\begin{equation}
L_X=2.4 \times 10^{39}~{\rm erg/s} \left(\frac{SFR}{1~M_\odot/yr}\right)~\text{at 0.5-2 keV}.
\end{equation}
HMXRB emit more significantly in the hard X-ray band than the soft
X-ray band. The IGM is optically thin to hard X-rays even at $z>10$,
thus a significant fraction of the X-ray photons are not available for
ionization or heating of the IGM until they get redshifted to lower
energies at $z \sim 2-3$ \citep{RicottiO:04, RicottiOG:05}.  The same
normal dwarf galaxies that host HMXRB have stars that radiate in the
\LW bands: for a SFR=1~M$_\odot$/yr the luminosity estimated using
Starburst~99 is $L_{LW}=3 \times 10^{43}$~erg/s (roughly in agreement
with the values given in \S~\ref{sec:disc}).

For the weak X-ray irradiation regime, $\beta_X<\beta_X^{\rm cr}$, the number
of \popIII stars depends only on the ratio $\beta_X$ of X-ray to the
\LW energy emitted by each source.  Assuming that normal galaxies dominate
the X-ray emission, the results are independent of the level of SFR in
early dwarf galaxies because both $L_X$ from HMXRB and $L_{LW}$ are
proportional to the SFR.  Thus, we find
\begin{equation}
\beta_X({\rm HMXRB})=10^{-4},
\end{equation}
that is lower than the contribution from normal SNe from \popIII stars, thus can be neglected. 

Our result is not in contradiction with simulation results from
\cite{Xu:14, Jeon:14, Jeon:15}. Those studies focus on local effects
(\ie, \popIII star formation and heating of the IGM in a small-box
simulation) produced by a pre-calculated X-ray background from HMXRBs
in low-metallicity dwarf galaxies (not coupled to star formation in the
box).  Our study instead focuses on global effects (\ie, in large
volumes) at high redshifts when the X-ray background can be calculated
self-consistently within a feedback loop in which the number of X-ray
sources producing the background is determined by the intensity of the
background they produce. However, our model is unable to capture local
effects and its validity starts to break down at redshifts $z<10$ when
halos with masses $>10^8$~M$_\odot$ and galaxies forming predominantly
\popII stars (due to metal pre-enrichment) become prevalent and
dominate the buildup of radiation backgrounds.
 
\subsection{Moving IMBHs in the ISM of primordial galaxies}

If \popIII stars produce IMBH remnants, these can accrete gas from the
ISM and radiate in normal dwarf galaxies after the minihalos hosting
\popIII stars merge to form more massive halos. The emission from IMBH
is thus proportional to the number of \popIII stars per unit volume as
required by our model. An IMBH of mass 100~M$_\odot$, moving at $v
\sim 20-30$~km/s in an ISM of density $10^4$~cm$^{-3}$, ISM typical
for early dwarf galaxies of mass $M_{dm}>10^7-10^8$~M$_\odot$, emits
continuously (without periodic luminosity burst typical of stationary
accreting BHs) at about $3\%$ of the Eddington rate: $L_X \approx 3
\times 10^{38}$~erg/s, $t_{on}\approx t_{H}(z=10)\approx 400$~Myr
\citep{ParkR:13}. Thus, if a fraction (in number), $f_{imbh}$, of the remnants of
\popIII stars is a $100$~M$_\odot$ IMBH accreting from the ISM at this
rate we find (assuming the fiducial $K_{LW}=10^{-2}$
  for $40$~M$_\odot$ \popIII stars):
\begin{equation}
\beta_X({\rm IMBH}) = 10^{-2} \left(\frac{f_{imbh}}{1.6\times 10^{-5}}\right).
\end{equation}

\subsection{Miniquasars}
Similarly to the case of IMBH, if a fraction of \popIII stars produce
massive BHs, the emission of miniquasars is proportional to the
number of \popIII stars, as required by the model.  Here we estimate
the contribution to the soft X-ray (mean) luminosity of the first
objects from the emission of rare miniquasars in the most massive
halos.  Let's assume as an example $10^5$~M$_\odot$ miniquasars
accreting at near the Eddington rate $L_{mqso}=10^{43}~{\rm
  erg/s}(M_{mqso}/10^5~M_\odot)$. For this mass BHs, based on
theoretical arguments \citep{ParkR:11, ParkR:12}, the period between
bursts is about $10$~Myr and the duty cycle $6\%$, also in agreement
with observations at lower redshift
\citep[\eg,][]{Steideletal:02}. Thus, during $t_{H}(z=10)\approx
400$~Myr the emission time is $t_{on}=24$~Myr. Assuming a fraction (in
number) of miniquasars per \popIII stars, $f_{mqso}$, and comparing to
the \LW luminosity from 40~M$_\odot$ \popIII stars, we get
\begin{equation}
\beta_X({\rm miniQSO})=10^{-2} \left(\frac{f_{mqso}}{8 \times 10^{-9}}\right)\left(\frac{M_{mqso}}{10^5~M_\odot}\right),
\end{equation}
meaning that one miniquasar of $10^5$~M$_\odot$ per $10^8$ \popIII
stars (or, assuming about 100 \popIII stars per cMpc$^3$, one miniquasar
per about $10^6$~cMpc$^3$) would boost the formation of \popIII stars
to near its maximum.

\section{Summary and Conclusions}\label{sec:sum}

A low level of X-ray emission in the early universe, although has a
negligible effect on reionization and the optical depth to Thompson
scattering, goes a long way in enhancing the number of \popIII stars
and dwarf galaxies with halo masses $M_{halo}<10^8$~M$_\odot$ that can
only form before IGM reheating and reionization at redshift $z \sim
6-10$. The maximum number of \popIII stars is obtained when the
critical halo mass in which gas can cool in less than a Hubble time
equals the Jeans mass of the IGM. This happens for $K_X=0.01
K_{LW}^{0.5}$, where $K_X$ and $K_{LW}$ are the mean energies of
the first sources of light in the soft X-ray and \LW bands in units of
$6\times 10^{53}$~ergs, respectively.

This low level of X-ray emission does not require assumptions on the
presence of unknown X-ray sources such as IMBH or miniquasars: it is
necessarily produced by SN explosions and SN remnants of \popIII
stars, thus is an unavoidable consequence of \popIII star formation
with a top heavy IMF (unless most stars are more massive than
$250$~M$_\odot$, that would lead to direct collapse into black holes
without SN explosion).  In addition, if a non-negligible fraction of
\popIII stars explode as hypernovae or PISNe, the soft X-rays from
their remnants and explosions is sufficiently large to promote the
formation of \popIII stars to about 400 per cMpc$^3$, that is near the
maximum number of \popIII stars with typical
masses $10-40$~M$_\odot$ that can form in any of our models with different $K_X$. A higher X-ray flux than the
one provided by \popIII hypernovae, for instance produced by accretion onto
IMBH from \popIII stars and miniquasars, would suppress the number
of \popIII stars because of the excessive heating of the IGM.  We find
that X-rays emitted by HMXRBs have a negligible effect in boosting the
number of \popIII stars when compared to the soft X-ray emission from
the first SN remnants.

The implications of a large number of \popIII stars include: i) a
copious production of black holes with masses similar to the ones
detected by LIGO ($\sim 10-30$~M$_\odot$) via gravitational wave
emission \citep{LIGO2016} (about $10^4$ BHs remnants of \popIII stars
are estimated within the Milky Way in the hypernova scenario); ii)
would provide supermassive black holes seeds; iii) although \popIII
stars cannot fully reionize the IGM due to their bursty nature, they
can contribute to the reionization process in a manner similar to an
early X-ray background \citep{HartleyRicotti2016}; iv) finally, since
the mean distance between minihalos hosting \popIII stars is small
($n_{pop3}^{-1/3} \sim 13$~kpc physical at $z \sim 10$), it is easier
for the metals ejected by their SN remnants to fill rather uniformly
the IGM. A low-level metal pre-enrichment of the IGM (\ie, the
metallicity floor often assumed in zoom simulations of galaxy
formation), promotes the formation of pre-reionization dwarf
galaxies and increases the number of their fossil relics in the Local
Group \citep{RicottiG:05, BovillR:09}. The population of ultra-faint
dwarfs discovered since 2005 orbiting the Milky Way
\citep{Belokurov2007a, DES2015, Koposov2015} is indeed consistent with
a large population of pre-reionization dwarf galaxies
\citep{BovillR:11a,BovillR:11b}.

\subsection*{ACKNOWLEDGMENTS}
I would like to thank the anonymous reviewer for the positive feedback and helping to improve 
the presentation of the material in the manuscript.
MR acknowledges support from NSF CDI-typeII grant CMMI1125285 and the
Theoretical and Computational Astrophysics Network (TCAN) grant
AST1333514.

\appendix

\section{Fits to Press-Schechter}

\begin{figure}
\includegraphics[width=9cm]{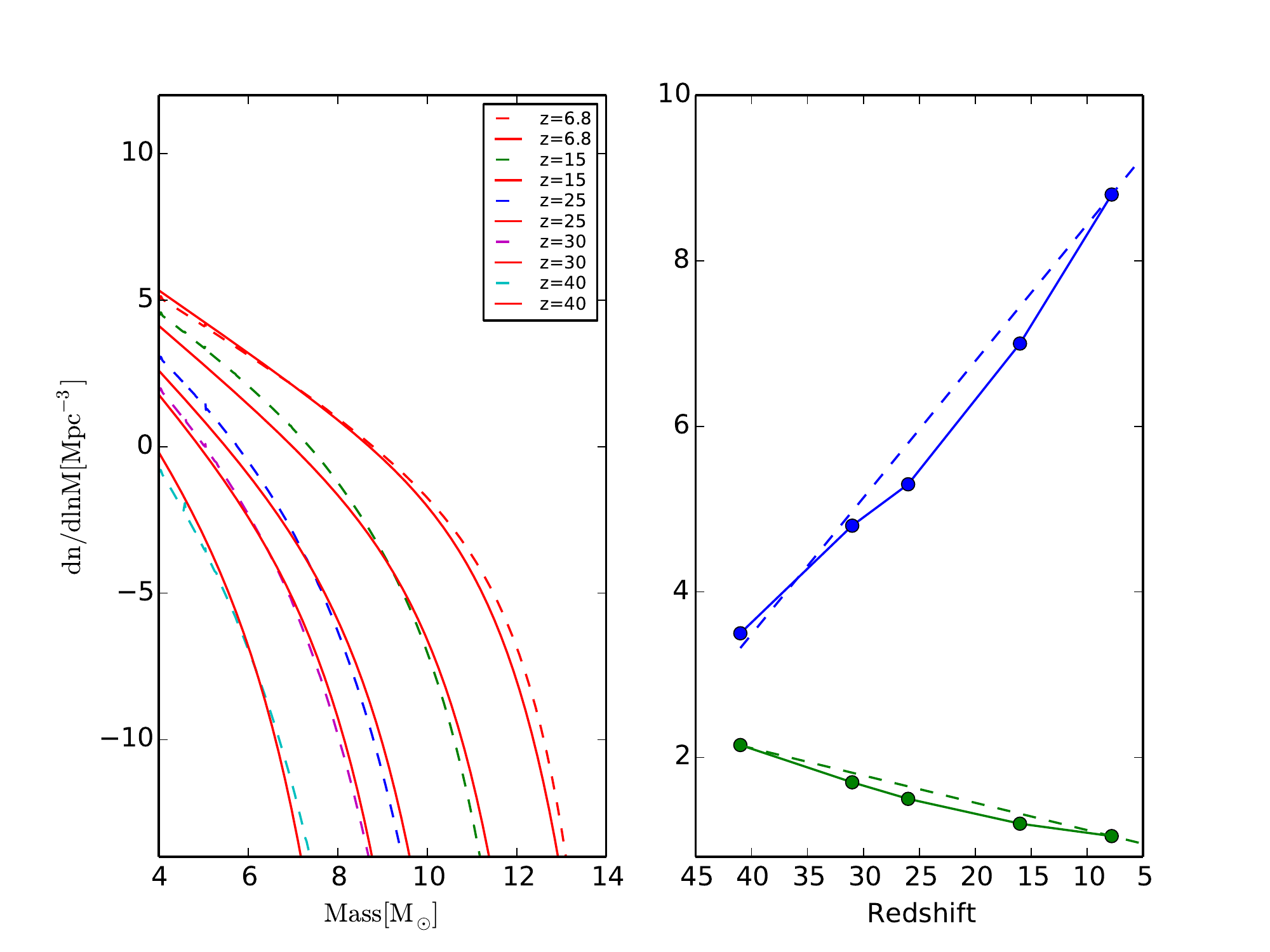}
\caption{\label{fig:ps} (Left) Mass function of dark matter halos
  from Press-Schechter formalism (solid lines) and fitting functions
  (dashed lines) from Equation~(\ref{eq:ps}). The redshifts (Right)
  Fitting parameters $\alpha$ (green line) and $M_*$ (blue line) as a function of redshift.}
\end{figure}

We use a modified Schechter equation to describe the dark matter mass
function of minihalos obtained from the Press-Schechter formalism
(including the Sheth-Tormen modification):
\begin{equation}
n_{\rm halo}(M,z)=3.98\times 10^{10}(M/0.1)^{-\alpha}\exp{[-(M/M_*)^{0.33}]},\label{eq:ps}
\end{equation}
where $n_{\rm halo}$ is the number of minihalos per unit $\ln{M}$ bin.  Fits
to the faint-end slope $\alpha=1.05+0.029(z1-7.8)$, and the truncation
mass $\log (M_*)=8.8-0.165(z1-7.8)$, are given as a function of
redshift $z$ for Planck cosmological parameters. In
Figure~\ref{fig:ps} are shown the the mass functions using
Press-Schechter formalism compared to the fitting function (left panel), and the fitting
parameters $\alpha$ and $M_*$ as a function of redshift.

\section{Rates}

In this appendix we summarize the rates used in our model. We adopt
the H$_2$ cooling rate $\Lambda(T)$ [erg cm$^3$ s$^{-1}$] from
\citep{Galli:98}:
\begin{align}
\log(\Lambda(T)) &= -103.0+97.59\log(T)-48.05\log(T)^2\\
&+10.80\log(T)^3-0.9032\log(T)^4
\end{align}

The H$^-$ formation rate and the H$_2$ dissociation rates are
respectively \citep{CazauxSpaans:04},
\begin{align}
k_1 & =1.4\times 10^{-18} T^{0.928}\exp(-T/16200.0)\\
k_2 & =\max(3.3\times 10^{-11} G_{LW}, t_H^{-1}).
\end{align}

The effect of secondary ionizations from photoelectrons is described by the following functions \citep{ShullVanS:85, ValdesFerrara:08}
\begin{align}
f_h &= 1 - 0.8751 ( 1 - x_e^{0.4})\\
f_i &= 0.3846( 1 - x_e^{0.54})^{1.19},
\end{align}
where $f_h$ is the fraction of fast photoelectrons that is thermalized
and $f_i$ the fraction that produce secondary ionizations.

%.......................................................................
%\bibliographystyle{/Users/ricotti/Latex/TeX/mn2e}
\bibliographystyle{/Users/ricotti/Latex/TeX/apj}
\bibliography{./first-background}

\label{lastpage}
\end{document}